\documentclass[aps,prd,twocolumn,nofootinbib]{revtex4-1}

\usepackage{graphicx}

\begin{document}

\title{Application of the Principle of Maximum Conformality to the Top Quark Forward-Backward Asymmetry at the Tevatron}

\author{Stanley J. Brodsky$^{1}$}
\email[email:]{sjbth@slac.stanford.edu}

\author{Xing-Gang Wu$^{1,2}$}
\email[email:]{wuxg@cqu.edu.cn}

\address{$^{1}$ SLAC National Accelerator Laboratory, 2575 Sand Hill Road, Menlo Park, CA 94025, USA\\
$^{2}$ Department of Physics, Chongqing University, Chongqing 401331, P.R. China}

\date{\today}

\begin{abstract}

The renormalization scale uncertainty can be eliminated by the Principle of Maximum Conformality (PMC) in a systematic scheme-independent way. Applying the PMC for the $t\bar{t}$-pair hadroproduction at the NNLO level, we have found that the total cross-sections $\sigma_{t\bar{t}}$ at both the Tevatron and LHC remain almost unchanged when taking very disparate initial scales $\mu^{\rm init}_R$ equal to $m_t$, $10\,m_t$, $20\,m_t$ and $\sqrt{s}$, which is consistent with renormalization group invariance. As an important new application, we apply PMC scale setting to study the top quark forward-backward asymmetry. We observe that the more convergent perturbative series after PMC scale setting leads to a more accurate top quark forward-backward asymmetry. The resulting PMC prediction on the asymmetry is also free from the initial renormalization scale-dependence. Because the NLO PMC scale has a dip behavior for the $(q\bar{q})$-channel at small subprocess collision energies, the importance of this channel to the asymmetry is increased. We observe that the asymmetries $A_{FB}^{t\bar{t}}$ and $A_{FB}^{p\bar{p}}$ at the Tevatron will be increased by $42\%$ in comparison to the previous estimates obtained by using conventional scale setting; i.e. we obtain $A_{FB}^{t\bar{t},{\rm PMC}} \simeq 12.5\%$ and $A_{FB}^{p\bar{p},{\rm PMC}} \simeq 8.28\%$. Moreover, we obtain $A_{FB}^{t\bar{t},{\rm PMC}}(M_{t\bar{t}}>450 \;{\rm GeV}) \simeq 35.0\%$. These predictions have a $1\sigma$-deviation from the present CDF and D0 measurements; the large discrepancies of the top quark forward-backward asymmetry between the Standard Model estimate and the CDF and D0 data are thus greatly reduced.  \\

\begin{description}

\item[PACS numbers] 12.38.Aw, 14.65.Ha, 11.15.Bt, 11.10.Gh
\item[Keywords] PMC, Renormalization Scale, top quark Forward-Backward Asymmetry

\end{description}

\end{abstract}

\maketitle

\section{Introduction}

The top quark is the heaviest known elementary particle, and it plays a fundamental role in testing the Standard Model (SM) and the extensions of the SM. Its production and decay channels are important probes of new physics, and because of its large coupling to the Higgs, the top quark production processes provide a sensitive probe of electroweak symmetry breaking. The total cross-section for the top quark pair production has been calculated up to NNLO within the $\overline{MS}$-scheme in Refs.~\cite{nason1,nason2,nason3,beenakker1,beenakker2,czakon1,moch1,moch2,hathor,moch3,czakon2,beneke1,beneke2,beneke3,andrea,vogt,pmc3,pmc4,alex,nik}. The SM estimates, especially those obtained by using the Principle of Maximum Conformality (PMC) \cite{pmc3,pmc4}, agree well with the experimental result which has been measured with a precision $\Delta \sigma_{t \bar{t}}/\sigma_{t \bar{t}}\sim\pm 7\%$ at the Tevatron~\cite{cdft,d0t} and $\sim \pm 10\%$ at the LHC~\cite{atlas,cms}.

The top quark forward-backward asymmetry which originates from charge asymmetry physics~\cite{cskim,Kuhn} has also been studied at the Tevatron and LHC. Two options for the asymmetry have been used for experimental analysis; i.e. the $t\bar{t}$-rest frame asymmetry
\begin{equation} \label{Afbtt}
A_{FB}^{t\bar{t}}= \frac{\sigma(y^{t\bar{t}}_t > 0) - \sigma(y^{t\bar{t}}_t < 0)}{\sigma(y^{t\bar{t}}_t > 0) + \sigma(y^{t\bar{t}}_t < 0)}
\end{equation}
and the $p\bar{p}$-laboratory frame asymmetry
\begin{equation} \label{Afblab}
A_{FB}^{p\bar{p}}= \frac{\sigma(y^{p\bar{p}}_{t} > 0) - \sigma(y^{p\bar{p}}_{t} < 0)}{\sigma(y^{p\bar{p}}_{t} > 0) + \sigma(y^{p\bar{p}}_{t} < 0)} ,
\end{equation}
where $y^{t\bar{t}}_t$ is the top quark rapidity in the $t\bar{t}$-rest frame and $y^{p\bar{p}}_{t}$ is the top quark rapidity in the $p\bar{p}$-laboratory frame (or the $p\bar{p}$ center-of-mass frame). The CDF and D0 collaborations have found comparable values in the $t\bar{t}$-rest frame: $A_{FB}^{t\bar{t},{\rm CDF}}=(15.8\pm7.5)\%$~\cite{cdf2} and $A_{FB}^{t\bar{t},{\rm D0}}=(19.6\pm6.5)\%$~\cite{d0}, where the uncertainties are derived from a combination of statistical and systematic errors. The asymmetry in the $p\bar{p}$-laboratory frame measured by CDF is $A_{FB}^{p\bar{p},{\rm CDF}} = (15.0\pm5.5)\%$~\cite{cdf2}. The CDF collaboration has also measured the dependence of $A^{t\bar{t}}_{FB}$ with respect to the $t\bar{t}$-invariant mass $M_{t\bar{t}}$: the asymmetry increases with $M_{t\bar{t}}$, and $A^{t\bar{t}}_{FB}(M_{t\bar{t}}>450\; {\rm GeV})=(47.5\pm11.4)\%$~\cite{cdf2}.

These measured top quark forward-backward asymmetries are much larger than the usual SM estimates. For example, the NLO QCD contributions to the asymmetric $t\bar{t}$-production using conventional scale setting yield $A_{FB}^{t\bar{t}}\simeq 7\%$ and $A^{p\bar{p}}_{FB}\simeq 5\%$ (see e.g.~\cite{zgsi}), which are about $2\sigma$-deviation from the above measurements. For the case of $M_{t\bar{t}}>450\; {\rm GeV}$, using the MCFM program~\cite{mcfm}, one obtains $A^{t\bar{t}}_{FB} (M_{t\bar{t}}>450\; {\rm GeV})\sim 8.8\%$ which is about $3.4\sigma$-deviation from the data. These discrepancies have aroused great interest because of the possibility for probing new physics beyond the Standard Model.

A recent reevaluation of the electroweak correction raises the QCD asymmetries by at most $20\%$: i.e. $A_{FB}^{t\bar{t}} (A^{p\bar{p}}_{FB}) \sim 9\%\;(7\%)$~\cite{qedc1,qedc2} and $A^{t\bar{t}}_{FB} (M_{t\bar{t}}>450\; {\rm GeV}) \sim 12.8\%$~\cite{qedc2}.

It has been argued that the missing higher-order corrections cannot be the reason for the significant discrepancy~\cite{nnloasy,nnloasy2,sterman,nki2}. In fact, only a several percent increment has been observed in Ref.\cite{nnloasy} using a next-to-next-to-leading-logarithmic (NNLL) calculation. It is for this reason that many new physics models beyond the SM have been suggested.

Since the SM estimate on the total cross-section agrees well with the experimental data, it is hard to understand the large deviation of the asymmetry. Before introducing any new physics, it is best to have a more precise estimation within the SM. It should be noted that all the present SM estimations are based on the conventional scale setting, where the renormalization scale $\mu_R$ is set to be the typical momentum transfer $Q$ of the process; i.e. $Q=m_t$. One then estimates the scale-uncertainty by varying $\mu_{R}\in[m_t/2,2m_t]$, which will lead to a ${\cal O}(10\%)$ scale uncertainty to the asymmetry. Usually, it is argued that this scale uncertainty can be suppressed by including higher-order corrections in an order-by-order manner. The conventional scale setting procedure is clearly problematic since the resulting fixed-order pQCD prediction will depend on the choice of renormalization scheme. In fact, it gives the wrong result when applied to QED processes.

It should be recalled that there is no ambiguity in setting the renormalization scale in QED. In the standard Gell-Mann-Low scheme for QED, the renormalization scale is the virtuality of the virtual photon~\cite{gml}. For example, the renormalization scale for the electron-muon elastic scattering through one-photon exchange can be set as the virtuality of the exchanged photon, i.e. $\mu^{GM-L}_R = Q =\sqrt{-q^2}$. But it is wrong to use $Q$ directly as the scale for any other renormalization scheme. Some displacement must be included in order to ensure scheme-independence. For example, under the $\overline {MS}$ scheme we have $\mu^{\overline {MS}}_R = e^{-5/6} Q\simeq 0.43Q$~\cite{Bardeen} \footnote{The same scale displacement can be obtained by using the PMC~\cite{pmc1,pmc2}. In fact, the PMC can also be applied to QED processes. One can obtain proper scale displacements among different renormalization schemes for higher perturbative orders in a systematic way. }. This result shows that the effective scale of the $\overline{MS}$ scheme should generally be about half of the true momentum transfer occurring in the interaction. The invariance under choice of renormalization scheme is a consequence of the transitivity property of the renormalization group~\cite{ren1,ren2}. Of course, the question is more complicated in QCD due to its non-Abelian nature.

Recently, it has been suggested that one can systematically fix the renormalization scale at any fixed order by using the PMC~\cite{pmc1,pmc2,pmc3,pmc4}. The PMC provides the principle underlying the Brodsky-Lepage-Mackenzie method~\cite{blm}, and they are consistent with each other through the PMC-BLM correspondence principle~\cite{pmc2}. The main idea is that, after proper procedures, all nonconformal $\{\beta_i\}$-terms in the perturbative expansion are summed into the running coupling so that the remaining terms in the perturbative series are identical to that of a conformal theory; i.e., the corresponding theory with $\{\beta_i\} \equiv \{0\}$. The underlying conformal symmetry is a useful principle for physics; e.g. the AdS/QCD theory~\cite{ads1}, the conformal general relativity model~\cite{cgr} and the canonical quantum gravity theory~\cite{thooft1}.

After PMC scale setting, the divergent renormalon series with $n!$-growth does not appear in the conformal series. This is consistent with the treatment done in Ref.\cite{kataev}. Since renormalon terms are absent, one obtains a more convergent perturbative expansion series, and thus the full next-to-leading order (NLO), or even the leading-order (LO) calculation, is often enough to achieve the required accuracy. The PMC scale $\mu^{\rm PMC}_R$ is unambiguous at any finite order. We emphasize that the PMC is consistent with the renormalization group property that a physical result is independent of the renormalization scheme and the choice of the initial renormalization scale $\mu^{\rm init}_R$. Any residual dependence on $\mu^{\rm init}_R$ for a finite-order calculation is highly suppressed since the unknown higher-order $\{\beta_i\}$-terms will be absorbed into the PMC scales' higher-order perturbative terms.

As an application, we have previously applied the PMC procedure to obtain NNLO predictions for the $t\bar{t}$-pair hadroproduction cross-section at the Tevatron and LHC colliders~\cite{pmc3,pmc4}. It is found that there is almost no dependence on the choice of initial renormalization scale; i.e. the total cross-section remains almost unchanged even when taking very disparate initial scales $\mu^{\rm init}_R$ equal to $m_t$, $10\,m_t$, $20\,m_t$ and $\sqrt{s}$, thus greatly improving the precision of the QCD prediction. By using the PMC scales, a larger $\sigma_{t\bar{t}}$ is obtained in comparison to the conventional scale setting, which agrees well with the present Tevatron and LHC data. It is thus interesting to see whether the use of PMC scales, especially those of the dominant asymmetric $(q\bar{q})$-channel at the Tevatron, can improve our understanding on the top quark forward-backward symmetry; this is the purpose of the present paper.

The remaining parts of this paper are organized as follows: in Sec.~II, we give the relevant formulae for the top quark forward-backward asymmetry. The new properties of the predictions after PMC scale setting are presented. In Sec.~III, we present the numerical results and some discussions for the top quark forward-backward asymmetry at the Tevatron. Sec.~IV provides a summary.

\section{The forward-backward asymmetry}

Before discussing the top quark forward-backward asymmetry, we first review the total cross-sections for the top quark pair production at the Tevatron up to NNLO.

A comparison of the production cross-sections using conventional scale setting versus PMC scale setting will give us some useful information on how the PMC scale setting can improve our understanding of the top quark pair production: the relative importance of all the production channels, especially those which provide the asymmetries; the convergence of the perturbative series for each production channel; etc. This information will be helpful for constructing a more precise perturbative expansion for calculating the top quark forward-backward asymmetry.

\begin{widetext}
\begin{center}
\begin{table}[ht]
\begin{tabular}{|c||c|c|c|c||c|c|c|c|}
\hline
& \multicolumn{4}{c||}{Conventional scale setting} & \multicolumn{4}{c|}{PMC scale setting} \\
\hline
~~~ ~~~    &~~~LO~~~  &~~~NLO~~~  &~~~NNLO~~~ &~~~ {\it total} ~~~&~~~LO~~~  &~~~NLO~~~  &~~~NNLO~~~ &~~~ {\it total} ~~~\\
\hline
$(q\bar{q})$-channel & 4.890 & 0.963 & 0.483 & 6.336 & 4.748 & 1.727 & -0.058 & 6.417 \\
\hline
$(gg)$-channel    & 0.526 & 0.440 & 0.166 & 1.132 & 0.524 & 0.525  & 0.160  & 1.208 \\
\hline
$(gq)$-channel    & 0.000 &-0.0381 & 0.0049& -0.0332 & 0.000 & -0.0381 & 0.0049  & -0.0332 \\
\hline
$(g\bar{q})$-channel & 0.000 &-0.0381 & 0.0049& -0.0332 & 0.000 & -0.0381 & 0.0049  & -0.0332 \\
\hline
sum     & 5.416  & 0.985 & 0.659 & 7.402 & 5.272 & 2.176  & 0.112  & 7.559 \\
\hline
\end{tabular}
\caption{Total cross-sections (in unit: pb) for the top quark pair production at the Tevatron with $p\bar{p}$-collision energy $\sqrt{S}=1.96$ TeV. For conventional scale setting, we set the renormalization scale $\mu_R\equiv Q$. For PMC scale setting, we set the initial renormalization scale $\mu^{\rm init}_R=Q$. Here we take $Q=m_t=172.9$ GeV and use the MSRT 2004-QED parton distributions~\cite{mrst} as the PDF. }\label{tab1}
\end{table}
\end{center}
\end{widetext}

Analytical expressions up to NNLO have been provided in the literature, e.g. Ref.\cite{moch1,moch2,beneke2,hathor} \footnote{These NNLO results are derived using resummation~\cite{moch1,moch2,beneke2}, which is supported by the observation that the production of a top quark pair with an additional jet is small~\cite{jet1,jet2}. The nearly scale-independent PMC estimations in Refs.~\cite{pmc3,pmc4} show that at least the relative importance of the $\{\beta_i\}$-terms at the NNLO have been well set. A full NNLO calculation for the $(q\bar{q})$-channel has been recently presented by using the conventional scale setting which also shows a very small perturbative uncertainty between the NNLO and NNLL calculation, i.e. $\pm2.7\%$~\cite{alex}. }, and the explicit calculation technology for PMC scale setting can be found in Ref.\cite{pmc4}, so we will not present them here. Numerical results for the top quark pair production at the Tevatron with $p\bar{p}$-collision energy $\sqrt{S}=1.96$ TeV are presented in Table \ref{tab1}. The Coulomb-type corrections will lead to sizable contributions in the threshold region~\cite{coul1,coul2} which are enhanced by factors of $\pi$. Thus the terms which are proportional to $(\pi/v)$ or $(\pi/v)^2$ ($v=\sqrt{1-4m_t^2/s}$ is the top quark velocity in the partonic center-of-mass frame; $s$ is the subprocess center-of-mass energy squared) at the NLO or NNLO level should be treated separately~\cite{brodsky1}. For this purpose, the results listed in the {\it total}-column is not a simple summation of the corresponding LO, NLO and NNLO results; the results are obtained by using the Sommerfeld rescattering formula to treat the Coulomb part. In doing the numerical calculation, we set $m_t=172.9$ GeV~\cite{pdg} and the factorization scale $\mu_f\equiv m_t$. We set $\mu_R\equiv Q=m_t$ for conventional scale setting, and take the initial renormalization scale $\mu^{\rm init}_R=Q=m_t$ to initialize PMC scale setting. For the PDFs, we adopt MSRT 2004-QED parton distributions~\cite{mrst} to be consistent with the choice of Ref.\cite{qedc2}.

By comparing with the total cross-sections derived from the PMC scale setting and the conventional scale setting listed in Table \ref{tab1}, we observe the following points:

\begin{itemize}

\item At the Tevatron, the top quark pair cross-section is dominated by the $(q\bar{q})$-channel which provides $\sim85\%$ contribution to the total cross-section. The $(q\bar{q})$-channel is asymmetric at the NLO level, so it will lead to sizable top quark forward-backward asymmetry at the Tevatron. In contrast, one finds that the dominant channel at the LHC is the symmetric $(gg)$-channel, c.f. Ref.\cite{Kuhn}, so the top quark forward-backward asymmetry from other channels will be greatly diluted at the LHC; this asymmetry becomes small which agrees with the CMS and ATLAS measurements~\cite{cms2,atlas2}. Accordingly, at present, we will concentrate on the top quark forward-backward asymmetry at the Tevatron.

\begin{figure}
\includegraphics[width=0.45\textwidth]{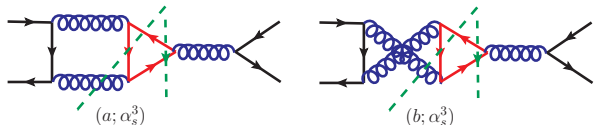}
\caption{Dominant Feynman diagrams (cut diagrams) for the QCD forward-backward asymmetry at the NLO level. Two types of asymmetries are shown: the interference of the final-state with the initial-state gluon bremsstrahlung and the interference of the box diagram with the Born diagram. }
\label{asy1}
\end{figure}

\begin{figure}
\includegraphics[width=0.25\textwidth]{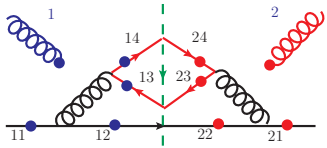}
\caption{Flavor-excitation Feynman diagrams (cut diagrams) for the QCD forward-backward asymmetry at the NLO level, which are small and negligible. Here either gluon-1 or gluon-2 can be attached at four places in the light quark or top quark lines, so there are totally 16 Feynman diagrams. }
\label{asy2}
\end{figure}

\item At the lowest order, the two channels $q\bar{q}\to t\bar{t}$ and $gg\to t\bar{t}$ do not discriminate the final top quark and top-antiquark, so their differential distributions are symmetric for the hadronic production process. At the NLO level, either the virtual or real gluon emission will cause sizable differences between the differential top quark and top-antiquark production, thus leading to an observable top quark forward-backward asymmetry. At the Tevatron, the asymmetric channels are $(q\bar{q})$-, $(gq)$- and $(g\bar{q})$- channels accordingly. Table \ref{tab1} shows the total cross-sections of the $(gq)$ and $(g\bar{q})$ channels are quite small, less than $1\%$ of that of $(q\bar{q})$-channel, so their contributions to the asymmetry can be safely neglected. Figure (\ref{asy1}) shows the dominant Feynman diagrams for the QCD charge asymmetry at the NLO level and Fig.(\ref{asy2}) shows the less important flavor-excitation Feynman diagrams for the QCD charge asymmetry at the NLO level.

\item When using conventional scale setting, the NNLO cross-section for the $(q\bar{q})$-channel is about $50\%$ of its NLO cross-section; i.e. $|\sigma^{\rm NNLO}_{t\bar{t},q\bar{q}}/\sigma^{\rm NLO}_{t\bar{t},q\bar{q}}| \simeq 50\%$. Thus in order to derive a consistent asymmetry up to NNLO, one must consider the asymmetric contribution from the NNLO $(q\bar{q})$-channel, which may be sizable. In contrast, a much more convergent pQCD series expansion is obtained after PMC scale setting, since all non-conformal $\{\beta_i\}$-terms in the perturbative expansion are summed into the running coupling. For example for the asymmetric $(q\bar{q})$-channel, the value of $|\sigma^{\rm NNLO}_{t\bar{t},q\bar{q}}/\sigma^{\rm NLO}_{t\bar{t},q\bar{q}}|$ is lowered to be only $\sim 3\%$. This shows that after PMC scale setting, the change to the asymmetry from the NNLO is greatly suppressed.

\item Writing the numerator and the denominator of the two asymmetries $A_{FB}$ defined by Eqs.(\ref{Afbtt},\ref{Afblab}) in powers of $\alpha_s$, we obtain
\begin{widetext}
\begin{eqnarray}
A_{FB} &=& \frac{\alpha_s^{3} N_{1}+\alpha_s^{4} N_{2}+ {\cal O}(\alpha_s^5)}{\alpha_s^{2} D_{0} +\alpha_s^{3} D_{1}+\alpha_s^{4} D_{2} +{\cal O}(\alpha_s^5)} \nonumber\\
&=& \frac{\alpha_s}{D_{0}}\left[N_{1}+ \alpha_s\left(N_{2}-\frac{D_{1} N_{1}}{D_{0}}\right)+\alpha_s^2 \left(\frac{D_1^2 N_1}{D_0^2} -\frac{D_1 N_2}{D_0} -\frac{D_2 N_1}{D_0}\right) +\cdots \right] ,
\end{eqnarray}
\end{widetext}
where the $D_i$-terms stand for the total cross-sections at certain $\alpha_s$-order and the $N_i$-terms stand for the asymmetric cross-sections at certain $\alpha_s$-order. The terms up to NLO ($D_{0},D_{1},N_{1}$) have been calculated, whereas only parts of $D_{2}$ and $N_{2}$ are currently known~\cite{nason1,nason2,nason3,beenakker1,beenakker2,czakon1,moch1,moch2,hathor,moch3,czakon2,beneke1,beneke2,beneke3,andrea,vogt}.

As shown in Table~\ref{tab1}, using conventional scale setting, the relative importance of the denominator terms is $\left[{\alpha_s^2} D_{0} : {\alpha_s^3} D_{1} : {\alpha_s^4}  D_{2} \simeq 1: 18\% : 12\% \right]$, and the numerator terms for the asymmetric $(q\bar{q})$-channel satisfy $\left[\alpha_s^{3} N_{1} : \alpha_s^{4} N_{2} \sim 1 : 50\% \right]$ \footnote{Since at present the NNLO numerator term $N_2$ is not available, as a first approximation, we treat these asymmetric terms to have the same relative importance as their total cross-sections; i.e. $(\alpha_s^{3} N_{1})_{q\bar{q}} : (\alpha_s^{4} N_{2})_{q\bar{q}} \sim (\alpha_s^{3} D_{1})_{q\bar{q}} : (\alpha_s^{4} D_{2})_{q\bar{q}}$. }. Thus, the $N_{1}D_{1}/D_{0}$ term and the $N_{2}$ term have the same importance. Because the NNLO $N_{2}$ term is not available at the present, one has to use the lowest-order $\mathcal{O}(\alpha_s^{2})$ cross-section in the denominator and the $\mathcal{O}(\alpha_s^{3})$ term in the numerator; i.e. dealing with only the so-called LO asymmetry~\cite{Kuhn,zgsi,qedc1,qedc2}: $A_{FB}=\frac{N_{1}}{D_{0}} \alpha_s$. \\

However, after PMC-scale setting, we have $\left[{\alpha_s^2} D_{0} : {\alpha_s^3} D_{1} : {\alpha_s^4}  D_{2} \simeq 1: 41\% : 2\%\right]$ and the numerators for the asymmetric $(q\bar{q})$-channel becomes $\left[\alpha_s^{3} N_{1} : \alpha_s^{4} N_{2} \sim 1 : 3\% \right]$. It shows that, after PMC scale setting, the NNLO corrections for both the total cross-sections and the asymmetric part are lowered by about one order of magnitude. Therefore, the NNLO-terms $N_2$ and $D_2$ can be safely neglected in the calculation, and we can obtain the asymmetry at the so-called NNLO level:
\begin{displaymath}
\quad\quad A_{FB}=\frac{\alpha_s}{D_{0}}\left[N_{1}-\alpha_s\left(\frac{D_{1} N_{1}}{D_{0}}\right)+\alpha_s^2 \left(\frac{D_1^2 N_1}{D_0^2}\right) \right] .
\end{displaymath}
Furthermore, it is natural to assume that those higher-order terms $N_i$ and $D_i$ with $i>2$ after PMC scale setting will also give negligible contribution \footnote{There may still be large higher-order corrections not associated with renormalization. The $n_f$-dependent but renormalization scale independent terms should not be absorbed into the coupling constant. An important example in QED case is the electron-loop light-by-light contribution to the sixth-order muon anomalous moment which is of order $(\alpha/\pi)^3\ln(m_{\mu}/m_{e})$~\cite{extraloop}. }; the above asymmetry can thus be resummed to a more convenient form:
\begin{equation}
A_{FB}= \frac{\alpha_s^{3} N_{1}}{\alpha_s^{2} D_{0} +\alpha_s^{3} D_{1}} \;.
\end{equation}

\begin{figure}
\includegraphics[width=0.48\textwidth]{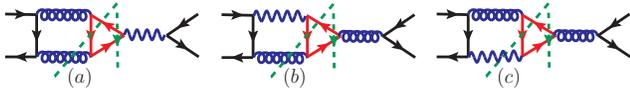}
\caption{Representative cut diagrams contributing to the QCD-QED interference term ${\cal O}(\alpha^2_s \alpha)$. The wave lines stand for the photon.}
\label{asyQED}
\end{figure}

\item As argued by Refs.~\cite{Kuhn,qedc1,qedc2}, the electromagnetic and weak interaction will provide an extra $\sim 20\%$ increment for the asymmetry. This shows that the electromagnetic contribution provides a non-negligible fraction of the QCD-based antisymmetric cross-section with the same overall sign. The asymmetry to be calculated thus changes to
\begin{equation}\label{final}
A_{FB}= \frac{\alpha_s^{3} N_{1}+\alpha_s^{2}\alpha \tilde{N}_{1}+\alpha^2 \tilde{N}_0}{\alpha_s^{2} D_{0} +\alpha_s^{3} D_{1}} \;.
\end{equation}
Representative diagrams contributing to the QCD-QED interference term $\tilde{N}_1$ at the order ${\cal O}(\alpha^2_s \alpha)$ are shown in Fig.(\ref{asyQED}). The weak contributions to the asymmetry are obtained by changing the photon propagator to be a $Z^0$-propagator. The pure electroweak antisymmetric ${\cal O}(\alpha^2)$ term $\tilde{N}_0$ arises from $|{\cal M}_{q\bar{q}\to\gamma\to t\bar{t}}+{\cal M}_{q\bar{q}\to Z^0\to t\bar{t}}|^2$~\cite{qedc2}.

\end{itemize}

\begin{figure}[ht]
\includegraphics[width=0.45\textwidth]{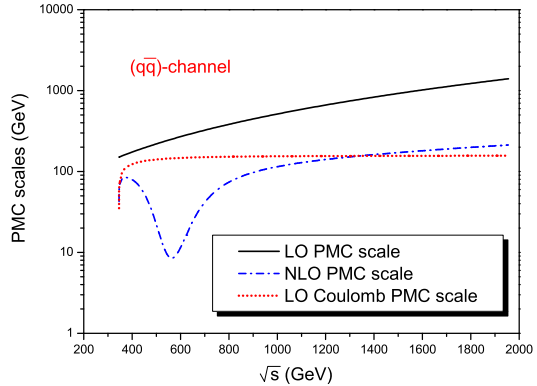}
\caption{PMC scales for the dominant asymmetry $(q\bar{q})$-channel versus the sub-process collision energy $\sqrt{s}$ for the top quark pair production up to $1.96$ TeV, where we have set the initial renormalization scale $\mu^{\rm init}_r=m_t=172.9$ GeV. }
\label{pmcscale}
\end{figure}

Based on the above considerations, the top quark forward-backward asymmetry after PMC scale setting can be written as
\begin{eqnarray}
A_{FB}^{t\bar{t},{\rm PMC}}&=& \frac{1} {\sigma^{\rm tot, PMC}_{H_1 H_2 \to t\bar{t}X}(\mu^{\rm PMC}_R)} \left[\sigma_{asy}^{(q\bar{q})}\left(\mu^{\rm PMC}_R; y^{t\bar{t}}_t > 0\right)\right. \nonumber\\
&& \left. \quad\quad\quad\quad\quad\quad\quad - \sigma_{asy}^{(q\bar{q})}\left(\mu^{\rm PMC}_R; y^{t\bar{t}}_t < 0\right) \right]
\end{eqnarray}
and
\begin{eqnarray}
A_{FB}^{p\bar{p},{\rm PMC}}&=& \frac{1}{\sigma^{\rm tot, PMC}_{H_1 H_2 \to t\bar{t}X}(\mu^{\rm PMC}_R)} \left[\sigma_{asy}^{(q\bar{q})}\left(\mu^{\rm PMC}_R; y^{p\bar{p}}_{t} > 0\right) \right.\nonumber\\
&& \left. \quad\quad\quad\quad\quad\quad\quad - \sigma_{asy}^{(q\bar{q})}\left(\mu^{\rm PMC}_R; y^{p\bar{p}}_{t} < 0\right) \right],
\end{eqnarray}
where $\sigma^{\rm tot}_{H_1 H_2 \to t\bar{t}X}$ is total hadronic cross-section up to NLO. The symbol $\sigma_{asy}^{(q\bar{q})}$ stands for the asymmetric cross-section of the $(q\bar{q})$-channel which includes the above mentioned ${\cal O}(\alpha_s^3)$, ${\cal O}(\alpha_s^2 \alpha)$ and ${\cal O}(\alpha^2)$ terms. Here $\mu^{\rm PMC}_R$ stands for the PMC scale. In the denominator for the total cross-section up to NLO, for each production channel, we need to introduce two LO PMC scales which are for the Coulomb part and non-Coulomb part accordingly, and one NLO PMC scale for the non-Coulomb part \footnote{Since the channels $(ij) = \{(q{\bar q}), (gg), (gq), (g\bar{q})\}$ are distinct and non-interfering, their PMC scales should be set separately~\cite{pmc4}.}. In the numerator, we only need the NLO PMC scale $\mu^{\rm PMC, NLO}_R$ for the $(q\bar{q})$-channel, since it is the only asymmetric component. Detailed processes for deriving these PMC scales can be found in Ref.\cite{pmc4}, which are obtained by using the cross-sections calculated within the $\overline{MS}$-scheme. We present the behaviors of the PMC scales for the dominant asymmetric $(q\bar{q})$-channel in Fig.(\ref{pmcscale}). Note that if the cross-sections are calculated within any other renormalization scheme, some proper scale-displacements to the present PMC scales will be automatically set by PMC scale setting so as to ensure the scheme-independence of the final estimation.

\begin{figure}
\includegraphics[width=0.40\textwidth]{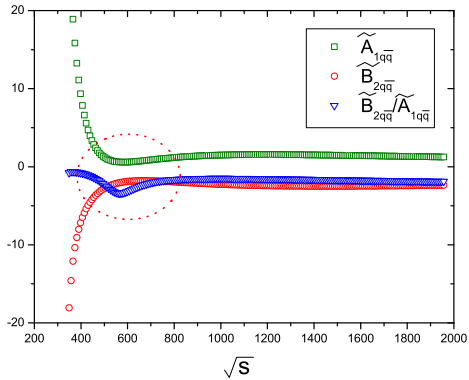}
\caption{PMC coefficients of the dominant asymmetric $(q\bar{q})$-channel versus the subprocess collision energy $\sqrt{s}$, which determine the dip behavior of the NLO PMC scale $\mu^{\rm PMC, NLO}_R$. $\mu^{\rm init}_R=m_t=172.9$ GeV. }
\label{qqcoe}
\end{figure}

It is interesting to observe that there is a dip for the NLO scale $\mu^{\rm PMC, NLO}_{R}$ of the $(q\bar{q})$-channel when $\sqrt{s} \simeq [\sqrt{2}\exp(5/6)]m_t \sim 563$ GeV, which is caused by the correlation among the PMC coefficients for NLO and NNLO terms. More specifically, it is found that
\begin{widetext}
\begin{eqnarray}
{\mu^{\rm PMC, NLO}_R} = \exp\left(\frac{\tilde{B}_{2q\bar{q}}} {\tilde{A}_{1q\bar{q}}}\right) {\mu^{\rm PMC, LO}_R}  =\exp\left(\frac{\tilde{B}_{2q\bar{q}}}{\tilde{A}_{1q\bar{q}}}\right) \exp\left(\frac{3B_{1q\bar{q}}}{2A_{0q\bar{q}}}+{\cal O}(\alpha_s) \right) {\mu^{\rm init}_R},
\end{eqnarray}
\end{widetext}
where the coefficients are defined through the standard PMC scale setting~\cite{pmc3,pmc4}; i.e.
\begin{widetext}
\begin{eqnarray}
[m_t^2 \hat\sigma_{q\bar{q}}]_{\rm non-Coulomb} &=& A_{0q\bar{q}} a^2_s({\mu^{\rm init}_R}) + \left[ A_{1q\bar{q}} + B_{1q\bar{q}} n_f  \right] a^3_s({\mu^{\rm init}_R}) +  \left[ A_{2q\bar{q}} + B_{2q\bar{q}} n_f + C_{2q\bar{q}} n^2_f \right] a^4_s({\mu^{\rm init}_R}) \nonumber\\
&=& A_{0q\bar{q}} a^2_s({\mu^{\rm PMC, LO}_R}) + \left[\tilde{A}_{1q\bar{q}}\right] a^3_s({\mu^{\rm PMC, LO}_R}) + \left[\tilde{A}_{2q\bar{q}} + \tilde{B}_{2q\bar{q}} n_f \right] a^4_s({\mu^{\rm PMC, LO}_R}) \nonumber\\
&=& A_{0q\bar{q}} a^2_s({\mu^{\rm PMC, LO}_R}) + \left[\tilde{A}_{1q\bar{q}}\right] a^3_s({\mu^{\rm PMC, NLO}_R}) + \left[\tilde{\tilde{A}}_{2q\bar{q}} \right] a^4_s({\mu^{\rm PMC, NLO}_R}) .
\end{eqnarray}
\end{widetext}
Here $\hat\sigma_{q\bar{q}}$ stands for the partonic cross-section. As shown in Fig.(\ref{qqcoe}), the value of $\tilde{B}_{2q\bar{q}}$ is always negative and $\tilde{A}_{1q\bar{q}}$ has a minimum value at small $\sqrt{s}$. As a result, there will be a dip for the NLO PMC scale $\mu^{\rm PMC, NLO}_R$ as shown in Fig.(\ref{pmcscale}). Quantitatively, the NLO PMC scale $\mu^{\rm PMC, NLO}_R$ for the $(q\bar{q})$-channel is considerably smaller than $m_t$ in the small $\sqrt{s}$-region (corresponding to small momentum fraction of the incident partons which are favored by the parton luminosity ${\cal L}_{q\bar{q}}$~\cite{pmc4}). The NLO cross-section of the $(q\bar{q})$-channel will thus be greatly increased; it is a factor of two times larger than its value derived under conventional scale setting, as shown by Table \ref{tab1}.

As a byproduct, it is found that if fixing the calculation only at the NLO level, i.e. there is no higher-order terms in the LO PMC scale \footnote{Note that the PMC scales will be a perturbative series of $\alpha_s$ so as to absorb all $n_f$-dependent terms properly~\cite{pmc2,stanlu}.}, and setting the initial renormalization scale to be equal to the factorization sale ${\mu^{\rm init}_R}=\mu_f\equiv m_t$, our present LO PMC scale ${\mu^{\rm PMC, LO}_R}$ for the $(q\bar{q})$-channel returns to the normal choice which agrees with the QED case,
\begin{displaymath}
{\mu^{\rm PMC, LO}_R} \cong \exp(-5/6)\sqrt{s} .
\end{displaymath}
Note if ${\mu^{\rm init}_R}\neq\mu_f$, one can apply the renormalization group method to derive the full scale-dependent coefficients~\cite{moch3} and then get the same result. The new terms which involve the factor $\ln \left({\mu^{\rm init}_R}^2/\mu^2_{f}\right)$ must be separated into two parts: one is proportional to $\ln\mu^2_f/m_t^2$ which should be kept in its original form, and the other one is proportional to $\ln\left({\mu^{\rm init}_R}^2/m_t^2\right)$ which should be absorbed into the lower-order $\alpha_s$-terms through the standard PMC scale setting.

\section{Phenomenological applications}

The PMC asymmetries $A_{FB}^{t\bar{t},{\rm PMC}}$ and $A_{FB}^{p\bar{p},{\rm PMC}}$ can be compared with the asymmetries calculated using conventional scale setting. For definiteness, we apply PMC scale setting to improve Hollik and Pagani's results~\cite{qedc2}, and we obtain
\begin{widetext}
\begin{eqnarray}  \label{pmcr1}
A_{FB}^{t\bar{t},{\rm PMC}}&=&  \left\{\frac{\sigma^{\rm tot, HP}_{H_1 H_2 \to t\bar{t}X} } {\sigma^{\rm tot, PMC}_{H_1 H_2 \to t\bar{t}X}} \right\} \left\{ \frac{{\overline{\alpha}_s}^3\left(\overline{\mu}^{\rm PMC, NLO}_R\right)} {{\alpha^{HP}_s}^3 \left(\mu^{\rm conv}_R\right)} A_{FB}^{t\bar{t},{\rm HP}}|_{{\cal O}(\alpha_s^3)} + \frac{{\overline{\alpha}_s}^2\left(\overline{\mu}^{\rm PMC, NLO}_R\right)} {{\alpha^{HP}_s}^2 \left(\mu^{\rm conv}_R\right)} A_{FB}^{t\bar{t},{\rm HP}}|_{{\cal O}(\alpha_s^2\alpha)}+ A_{FB}^{t\bar{t},{\rm HP}}|_{{\cal O}(\alpha^2)} \right\}\\
A_{FB}^{p\bar{p},{\rm PMC}}&=&  \left\{\frac{\sigma^{\rm tot, HP}_{H_1 H_2 \to t\bar{t}X} } {\sigma^{\rm tot, PMC}_{H_1 H_2 \to t\bar{t}X}} \right\} \left\{ \frac{{\overline{\alpha}_s}^3\left(\overline{\mu}^{\rm PMC, NLO}_R\right)} {{\alpha^{HP}_s}^3 \left(\mu^{\rm conv}_R\right)} A_{FB}^{p\bar{p},{\rm HP}}|_{{\cal O}(\alpha_s^3)} + \frac{{\overline{\alpha}_s}^2\left(\overline{\mu}^{\rm PMC, NLO}_R\right)} {{\alpha^{HP}_s}^2 \left(\mu^{\rm conv}_R\right)} A_{FB}^{p\bar{p},{\rm HP}}|_{{\cal O}(\alpha_s^2\alpha)}+ A_{FB}^{p\bar{p},{\rm HP}}|_{{\cal O}(\alpha^2)} \right\}  \label{pmcr2}
\end{eqnarray}
\end{widetext}
Here $\mu^{\rm conv}_R$ stands for the renormalization scale set by conventional scale setting and the symbol HP stands for the corresponding values of Ref.\cite{qedc2}; i.e. for $\mu^{\rm conv}_R =m_t$, it shows~\cite{qedc2}
\begin{eqnarray}
&& \sigma^{\rm tot, HP}_{H_1 H_2 \to t\bar{t}X} = 5.621 \;{\rm pb} \nonumber \\
&& A_{FB}^{t\bar{t},{\rm HP}}|_{{\cal O}(\alpha_s^3)}=7.32\% \;\;\; A_{FB}^{t\bar{t},{\rm HP}}|_{{\cal O}(\alpha_s^2\alpha)}=1.36\% \nonumber\\
&& A_{FB}^{t\bar{t},{\rm HP}}|_{{\cal O}(\alpha^2)}=0.26\% \;\;\; A_{FB}^{p\bar{p},{\rm HP}}|_{{\cal O}(\alpha_s^3)}=4.85\% \nonumber \\
&& A_{FB}^{p\bar{p},{\rm HP}}|_{{\cal O}(\alpha_s^2\alpha)}=0.90\% \;\;\; A_{FB}^{p\bar{p},{\rm HP}}|_{{\cal O}(\alpha^2)}=0.16\%  \nonumber
\end{eqnarray}
where
\begin{itemize}
\item $A_{FB}^{t\bar{t},{\rm HP}}|_{{\cal O}(\alpha_s^3)}$ and $A_{FB}^{p\bar{p},{\rm HP}}|_{{\cal O}(\alpha_s^3)}$ stand for the pure QCD asymmetry at the $\alpha_s^3$-order under the $t\bar{t}$-rest frame and the $p\bar{p}$ lab frame respectively.
\item $A_{FB}^{t\bar{t},{\rm HP}}|_{{\cal O}(\alpha_s^2\alpha)}$ and $A_{FB}^{p\bar{p},{\rm HP}}|_{{\cal O}(\alpha_s^2\alpha)}$ stand for the combined QED and weak with the QCD asymmetry at the $\alpha_s^2 \alpha$-order under the $t\bar{t}$-rest frame and the $p\bar{p}$ lab frame respectively.
\item $A_{FB}^{t\bar{t},{\rm HP}}|_{{\cal O}(\alpha^2)}$ and $A_{FB}^{p\bar{p},{\rm HP}}|_{{\cal O}(\alpha^2)}$ stand for the pure electroweak asymmetry at the $\alpha^2$-order under the $t\bar{t}$-rest frame and the $p\bar{p}$ lab frame respectively.
\end{itemize}

\begin{figure}[ht]
\includegraphics[width=0.45\textwidth]{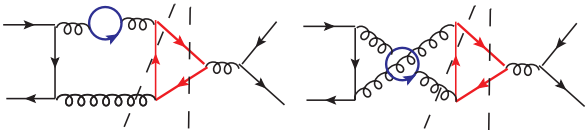}
\caption{Dominant cut diagrams for the $n_f$-terms at the $\alpha^4$-order of the $(q\bar{q})$-channel, which are responsible for the smaller effective NLO PMC scale $\overline{\mu}^{\rm PMC, NLO}_R$, where the solid circles stand for the light quark loops. }
\label{nloscale}
\end{figure}

In the formulae (\ref{pmcr1},\ref{pmcr2}), we have defined an effective coupling constant ${\overline{\alpha}_s} \left(\overline{\mu}^{\rm PMC, NLO}_R \right)$ for the asymmetric part, which is the weighted average of the strong coupling constant for the $(q\bar{q})$-channel; i.e. in using the effective coupling constant ${\overline{\alpha}_s} \left(\overline{\mu}^{\rm PMC, NLO}_R \right)$, one obtains the same $(q\bar{q})$-channel NLO cross-section as that of ${\alpha}_s (\mu^{\rm PMC, NLO}_R )$\footnote{This mean value technology is consistent with the global PMC scale idea suggested in Ref.\cite{pmc1}. In principle, one could divide the cross-sections into the symmetric and asymmetric components and to find PMC scales for each of them. For this purpose, one needs to identify the $n_f$-terms or the $n_f^2$-terms for both the symmetric and asymmetric parts at the NNLO level separately. }. It is noted that the NLO-level asymmetric part for $(q\bar{q})$-channel only involves the NLO PMC scale for the non-Coulomb part, so the effective coupling constant ${\overline{\alpha}_s}\left(\overline{\mu}^{\rm PMC, NLO}_R\right)$ can be unambiguously determined. We obtain a smaller effective NLO PMC scale
\begin{equation}
\overline{\mu}^{\rm PMC, effective}_R \simeq \exp(-9/10)m_t\sim 70 \;{\rm GeV} \;,
\end{equation}
which corresponds to
\begin{equation} \label{pmcsc}
{\overline{\alpha}_s}\left(\overline{\mu}^{\rm PMC, NLO}_R\right)=0.1228 .
\end{equation}
It is larger than ${\alpha^{HP}_s}\left(m_t\right)\simeq 0.098$~\cite{qedc1,qedc2}.
This effective NLO PMC scale is dominated by the non-Coulomb $n_f$-terms at the $\alpha^4$-order, which are shown in Fig.(\ref{nloscale}). In these diagrams, the momentum flow in the virtual gluons possess a large range of virtualities. This effect for NLO PMC scale $\overline{\mu}^{\rm PMC, effective}_R$ can be regarded as a weighted average of these different momentum flows in the gluons, so it can be small.

\begin{figure}
\includegraphics[width=0.45\textwidth]{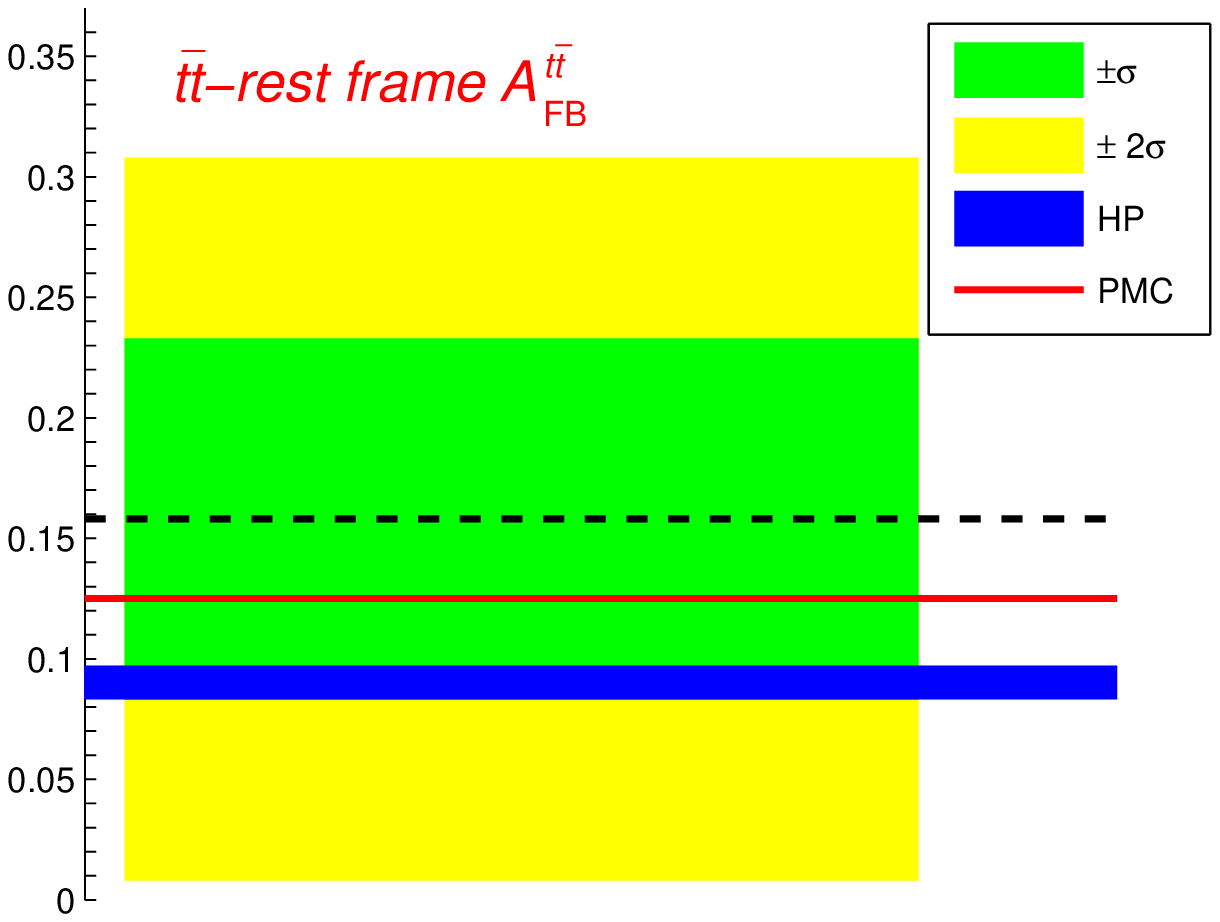}
\includegraphics[width=0.45\textwidth]{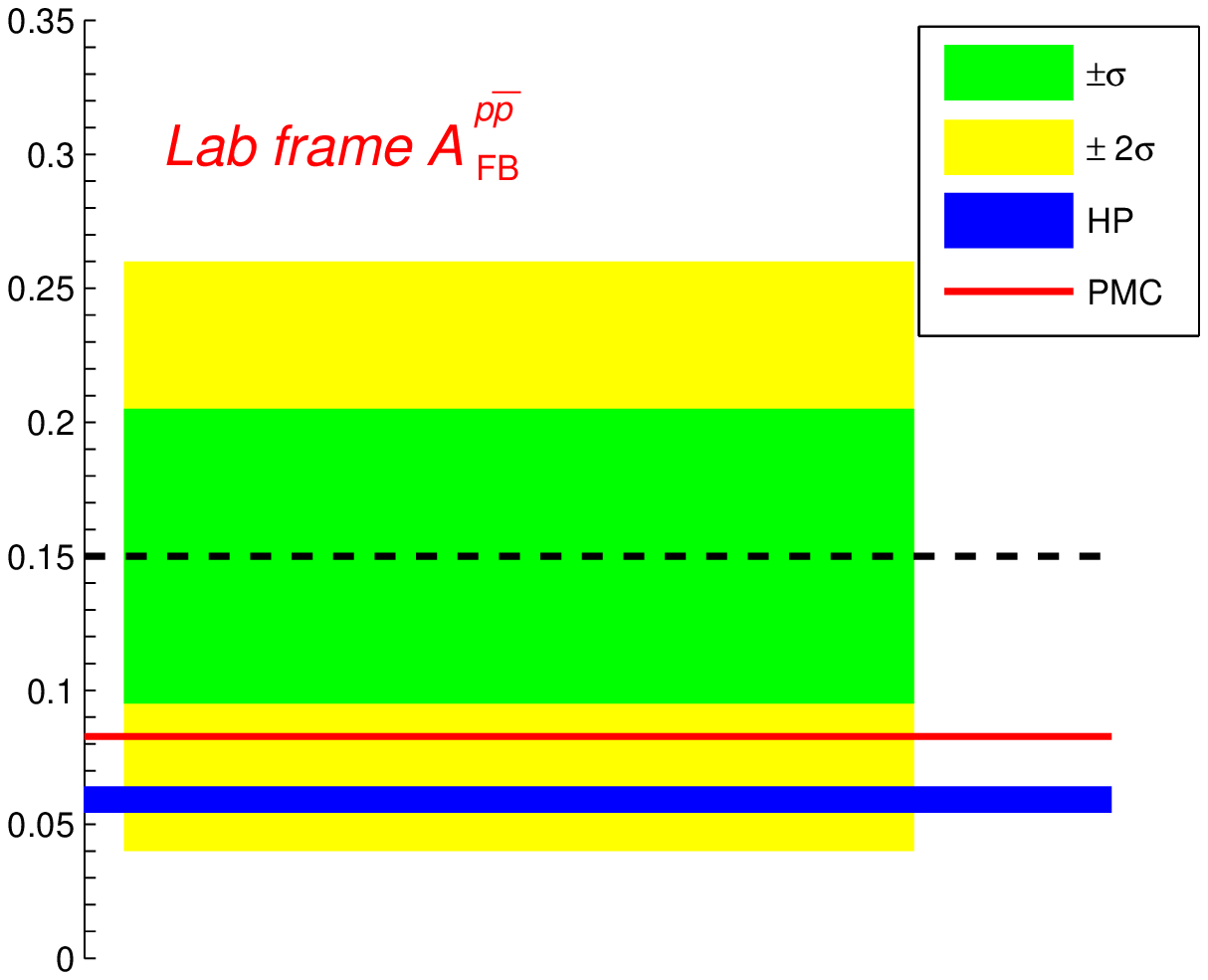}
\caption{Comparison of the PMC prediction with the CDF data~\cite{cdf2} for the $t\bar{t}$-pair forward-backward asymmetry for the whole phase space. The upper diagram is for $A_{FB}^{t\bar{t}}$ in the $t\bar{t}$-rest frame and the lower diagram is for $A_{FB}^{p\bar{p}}$ in the laboratory frame. The Hollik and Pagani's results (HP)~\cite{qedc2} using conventional scale setting are presented for a comparison. The result for D0 data~\cite{d0} shows a similar behavior. } \label{pmcasy}
\end{figure}

Finally, we obtain
\begin{equation}
A_{FB}^{t\bar{t},{\rm PMC}} \simeq 12.7\% \; ;\;\; A_{FB}^{p\bar{p},{\rm PMC}} \simeq 8.39\%
\end{equation}
Thus, after PMC scale setting, the top quark asymmetry under the conventional scale setting is increased by $\sim 42\%$ for both the $t\bar{t}$-rest frame and the $p\bar{p}$-laboratory frame. This large improvement is explicitly shown in Fig.(\ref{pmcasy}), where Hollik and Pagani's results which are derived under conventional scale setting~\cite{qedc2} are presented for comparison. In Fig.(\ref{pmcasy}), the upper diagram is for $A_{FB}^{t\bar{t}}$ in the $t\bar{t}$-rest frame and the lower diagram is for $A_{FB}^{p\bar{p}}$ in the laboratory frame.

\subsection{The PMC prediction of $A_{FB}^{t\bar{t}}(M_{t\bar{t}}>450\; {\rm GeV})$}

\begin{figure}
\includegraphics[width=0.45\textwidth]{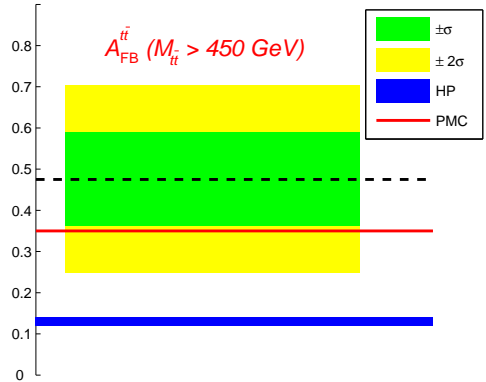}
\caption{The PMC prediction of $A_{FB}^{t\bar{t}}(M_{t\bar{t}}>450\; {\rm GeV})$ and the corresponding CDF data~\cite{cdf2} for the $t\bar{t}$-pair forward-backward asymmetry for $M_{t\bar{t}}>450$ GeV. The Hollik and Pagani's results (HP)~\cite{qedc2} using conventional scale setting are presented for a comparison. } \label{asycut}
\end{figure}

The CDF collaboration has found that when the $t\bar{t}$-invariant mass, $M_{t\bar{t}}>450$ GeV, the top quark forward-backward asymmetry $A^{t\bar{t}}_{FB}(M_{t\bar{t}}>450 \;{\rm GeV})$ is about $3.4$ standard deviations above the SM asymmetry prediction under the conventional scale setting~\cite{mcfm}. However, after applying PMC scale setting, with the help of the formulae (\ref{pmcr1},\ref{pmcr2}) and the cross-sections derived by using conventional scale setting which are listed in Ref.\cite{qedc2}, we will obtain a much larger $A^{t\bar{t}}_{FB}(M_{t\bar{t}}> 450 \;{\rm GeV})$ than the previous estimation~\cite{qedc2}.

For the present case \footnote{For simplicity, we have adopted the partonic center-of-mass frame to estimate the PMC total cross-section under the condition of $M_{t\bar{t}}>450\; {\rm GeV}$, which however agrees with that of $t\bar{t}$-rest frame within a high accuracy, since as shown in Ref.\cite{nnloasy}, the events near the partonic threshold provide the dominant contributions to the cross-section at the Tevatron.}, we have $\sigma^{\rm tot, PMC}_{H_1 H_2 \to t\bar{t}X}(M_{t\bar{t}}>450\; {\rm GeV})=2.406$ pb and $${\overline{\alpha}_s}\left(\overline{\mu}^{\rm PMC, NLO}_R\right)=0.1460 $$ with $$\overline{\mu}^{\rm PMC, NLO}_R \sim\exp(-19/10)m_t \simeq 26\;{\rm GeV}. $$ Then, we obtain
\begin{equation}
 A^{t\bar{t},PMC}_{FB}(M_{t\bar{t}}>450\; {\rm GeV})\simeq 35.0\% \;,
\end{equation}
which is increased by about $1.7$ times of the previous one $A_{FB}^{t\bar{t},{\rm HP}}(M_{t\bar{t}}>450\; {\rm GeV})=12.8\%$~\cite{qedc2}. Our present prediction is only about $1\sigma$-deviation from the CDF data, which is shown in Fig.(\ref{asycut}).

\subsection{Initial renormalization scale dependence}

We emphasize that the top quark asymmetry calculated under PMC scale setting is almost free of renormalization scale dependence.

\begin{figure}[ht]
\includegraphics[width=0.45\textwidth]{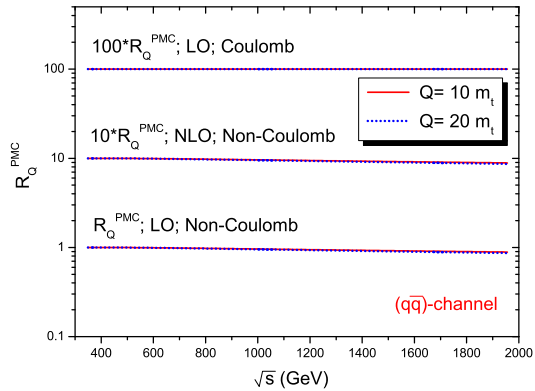}
\caption{The ratio $R_Q^{\rm PMC}= \frac{\mu^{\rm PMC}_{r}|_{\mu^{\rm init}_{r}=Q}}{\mu^{\rm PMC}_{r}|_{\mu^{\rm init}_r = m_t} } $ versus the sub-process collision energy $\sqrt{s}$ up to $1.96$ TeV for the $(q\bar{q})$-channel, where $Q=10\,m_t$ and $20\,m_t$ respectively. Here $m_t=172.9$ GeV. }
\label{ratio}
\end{figure}

To show how the change of initial scale affects the PMC scales, we define the ratio
\begin{displaymath}
R_Q^{\rm PMC}=\frac{\mu^{\rm PMC}_R|_{\mu^{\rm init}_{R}=Q}}{\mu^{\rm PMC}_R|_{\mu^{\rm init}_{R}=m_t}} ,
\end{displaymath}
where $\mu^{\rm PMC}_R|_{\mu^{\rm init}_{R}=Q}$ stands for the PMC scales determined under the condition of $\mu^{\rm init}_{R}=Q$. In Fig.(\ref{ratio}), we show the ratio $R_Q^{\rm PMC}$ versus the sub-process collision energy $\sqrt{s}$ up to $1.96$ TeV for the $(q\bar{q})$-channel, where $Q=10\,m_t$ and $20\,m_t$ respectively. The residual scale dependence for the PMC scales slightly increases with the subprocess collision energy $\sqrt{s}$; i.e. for the interested non-Coulomb NLO PMC scale, the value of $R_Q^{\rm PMC}$ is about $11\%$ for $Q=10\,m_t$ and $13\%$ for $Q=20\;m_t$ at $\sqrt{s}=1.96$ TeV. The cross-section at high collision energies is strongly suppressed by the parton luminosities, so that the total cross-section at the Tevatron remains almost unchanged even when taking disparate initial scales $\mu^{\rm init}_R$ equal to $m_t$, $10\,m_t$, $20\,m_t$. Due to this fact, the top quark asymmetry after PMC scale setting is also almost free of initial renormalization scale dependence; i.e. the residual scale uncertainty is less than $10^{-3}$ by taking $Q=m_t/4$, $10\, m_t$, $20 \, m_t$ and $\sqrt{s}$ respectively. \\

\section{summary}

With the help of present known top quark pair production cross-sections up to NNLO, we have presented a new analysis on the top quark forward-backward asymmetry using PMC scale setting. After PMC scale setting, a more convergent pQCD series expansion is obtained and the renormalization scale and scheme ambiguities are removed.

In comparison to the previous SM values estimated under conventional scale setting, we have shown that after PMC scale setting, both the top quark forward-backward asymmetries $A_{FB}^{t\bar{t}}$ and $A_{FB}^{p\bar{p}}$ for $t\bar{t}$-rest frame and $p\bar{p}$-laboratory frame can be increased by $\sim 42\%$; i.e.
\begin{displaymath}
A_{FB}^{t\bar{t},{\rm PMC}} \simeq 12.7\% \;\;{\rm and}\;\; A_{FB}^{p\bar{p},{\rm PMC}} \simeq 8.39\% \;.
\end{displaymath}
Moreover, the top quark asymmetry with certain kinematical cut, such as $A_{FB}^{t\bar{t}}(M_{t\bar{t}}>450 \;{\rm GeV})$, can be raised by about $1.7$ times; i.e.
\begin{displaymath}
A^{t\bar{t},PMC}_{FB}(M_{t\bar{t}}>450 \;{\rm GeV})\simeq 35.0\%  \;.
\end{displaymath}
This shows that, after PMC scale setting, the top quark forward-backward asymmetries are close to the CDF and D0 measurements within only $\sim 1\sigma$-deviation. The discrepancies between the SM estimate and the present CDF and D0 data are greatly reduced. This greatly suppresses the parameter space for new physics.

It is clear that the previous large discrepancy between the SM estimation and the CDF and D0 data for the top quark forward-backward asymmetry is caused by the improper setting of the renormalization scale. The PMC provides a systematic way to obtain optimal renormalization scales for the high energy process, whose theoretical predictions are essentially free of initial renormalization scale dependence even at fixed order. As we have shown the top quark pair total cross-section and its forward-backward asymmetry are almost unaltered by taking very disparate initial renormalization scales at the NNLO level.

\hspace{1cm}

{\bf Acknowledgements}: We thank Leonardo di Giustino, Benedict von Harling, Alexander Mitov, Michal Czakon and Andrei Kataev for helpful conversations. This work was supported in part by the Program for New Century Excellent Talents in University under Grant NO.NCET-10-0882, Natural Science Foundation of China under Grant NO.11075225, and the Department of Energy contract DE-AC02-76SF00515. SLAC-PUB-15006.

\end{document}